# Phase transitions of the dimerized Kane-Mele model with/without the strong interaction

Tao Du[1*], Yue-Xun Li[1], Yan Li[2], He-Lin Lu[1], Hui Zhang[1]

[1]Department of Physics, Yunnan Minzu University, Kunming 650500, China

[2]Department of Electronic Information, Yunnan Minzu University, Kunming 650500, China

[*]E-mail: dutao@ymu.edu.cn

**Abstract.** The dimerized Kane-Mele model with/without the strong interaction is studied using analytical methods. The boundary of the topological phase transition of the model without strong interaction is obtained. Our results show that the occurrence of the transition only dependent on dimerized parameter $\alpha$ . From the one-particle spectrum, we obtain the completed phase diagram including other phases. Then, using different mean-field methods, we investigate the Mott transition and the magnetic transition of the strongly correlated dimerized Kane-Mele model. In the region between the two transitions, the topological Mott insulator (TMI) with characters of Mott insulators and topological phases may be the most interesting phase. In this work, effects of the hopping anisotropy and Hubbard interaction U on boundaries of the two transitions are observed in detail. The completed phase diagram of the dimerized Kane-Mele-Hubbard model is also obtained in this work. Quantum fluctuations have extremely important influences on a quantum system. However, investigations are under the framework of the mean field treatment in this work and the effects of fluctuations in this model will be discussed in the future.

## 1. Introduction

Over the past decade, topological insulators have been a main topic in condensed matter physics [1-6]. Among them, the quantum spin Hall (QSH) state, or the two-dimensional time reversal invariant topological insulator attracted a lot of attentions, which was investigated originally by Kane and Mele [7, 8] in 2005. The novel quantum phase results from its nontrivial band topology induced by spin-orbit interaction, and is characterized by a $Z_2$ topological invariant. In fact, it can be regarded as two copies of the quantum anomalous Hall (QAH) state for each spin sector. The QAH state was investigated originally by Haldane [9] in 1988 and is characterized by a topological invariant named Chern number [10]. The Kane-Mele model possessing QSH state is a significant toy model in studies of topological insulators. Bernevig, Hughes and Zhang [11] predicted a QSH state in the HgTe/CdTe quantum well. The first experimental confirmation of the existence of the QSH state in the HgTe/CdTe quantum well was carried out by König *et al* [12] in 2007. So far, there are extensive studies on topological insulators in several systems without strongly interactions.

Actually, weak disorder or many-body interactions do not destroy topological phases due to the topological nature of phases [2, 13]. When the topology and electron correlations are both substantial, in several cases, they are competing each other. The topological Mott insulator [14] was suggested to describe the novel quantum state which has characteristics of topological band insulators and Mott insulators. Other novel quantum states stemming from the interplay of topology and electron correlations such as fractionalized Chern insulator (CI) [15, 16], fractionalized QSH [17, 18], etc. have been attracted extensive attentions in recent years. There are a large number of investigations of strongly correlated effects on topological insulators using several analytical or numerical methods, e.g. slave-particles mean field theory [19-22], quantum Monte Carlo simulations (QMC) [23-25], cellular dynamical mean field theory (CDMFT) [26], and variational cluster approach (VCA) [27].

Correlated topological insulators contain two classes: interaction-driven topological insulators [28] and systems with strong spin-orbit coupling to which interactions are introduced. In this work, we investigate the dimerized Kane-Mele model and introduce the Hubbard interaction to this model to analyze effects of the lattice anisotropy and the strong interaction using slave-rotor mean field theory [29, 30]. The rest of this paper is organized as follows. In section 2 we study the dimerized Kane-Mele model and obtain the phase diagram of this system. In section 3 the interaction is introduced in the anisotropic lattice system and we investigated in detail two scenarios of the phase transition. Finally, we conclude in section 4.

## 2. The dimerized Kane-Mele model without interactions

The generalized Kane-Mele model on the honeycomb lattice is

$$H = -\sum_{<i,j>} t_{ij} \hat{c}_{i\sigma}^{\dagger} \hat{c}_{j\sigma} + \sum_{<<i,j>>} \sum_{\sigma\sigma'} i\lambda_{ij} \nu_{ij} \hat{c}_{i\sigma}^{\dagger} \sigma_{\sigma\sigma'}^{z} \hat{c}_{j\sigma'} . \tag{1}$$

Where $\hat{c}_{i\sigma}^{\dagger}$ ($\hat{c}_{i\sigma}$) is a creation (annihilation) operator for an electron of spin $\sigma = \pm 1$ at site $i$. $\sigma_{\sigma\sigma'}^{z}$ is the z component of Pauli matrices, $t_{ij}$ is the hopping amplitude of electrons at nearest neighbor sites (NN) and $\lambda_{ij}$ is the strength of spin-orbital coupling of electrons at next neighbor sites (NNN). In the so called dimerized Kane-Mele model, $t_{ij} = \alpha t$ when sites $i, j$ are endpoints of the bond $\vec{\delta}_1$, $t_{ij} = t$ for other bonds (Fig.1) and $\lambda_{ij} = \lambda$ for all of NNN sites. The parameter $\nu_{ij} = +1$ when the orientation of the NNN sites $i, j$ is left turn while $\nu_{ij} = -1$ when right turn. The honeycomb lattice comprises of two sublattice A and B respectively and lattice vectors are $\vec{a}_1 = (3a/2, \sqrt{3}a/2)$ and $\vec{a}_2 = (3a/2, -\sqrt{3}a/2)$ as shown in Fig.1.The Hamiltonian of the dimerized Kane-Mele model is

$$H_{DKM} = -\alpha t \sum_{<i,j>\delta_1} \hat{c}_{i\sigma}^{\dagger} \hat{c}_{j\sigma} + t \sum_{<i,j>\delta_{2,3}} \hat{c}_{i\sigma}^{\dagger} \hat{c}_{j\sigma} + i\lambda \sum_{<<i,j>>} \sum_{\sigma\sigma'} \nu_{ij} \hat{c}_{i\sigma}^{\dagger} \sigma_{\sigma\sigma'}^{z} \hat{c}_{j\sigma'} . \tag{2}$$

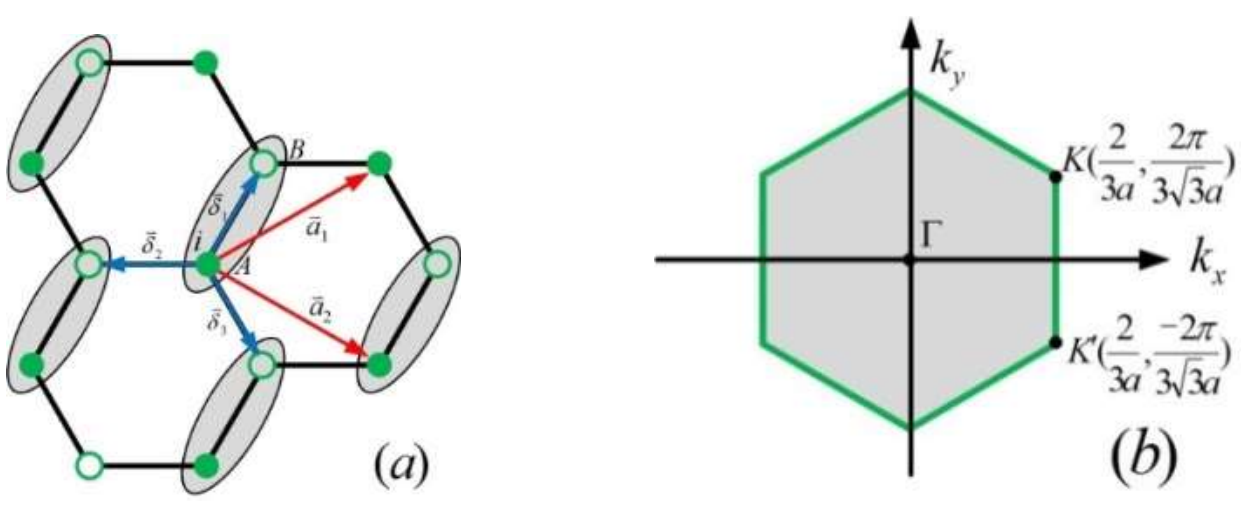

Fig.1. (a) The dimerized honeycomb lattice. Red arrows represent the lattice vectors $\vec{a}_1 = (3a/2, \sqrt{3}a/2)$ and $\vec{a}_2 = (3a/2, -\sqrt{3}a/2)$. Blue arrows represent bonds in three directions: $\vec{\delta}_1 = (a/2, \sqrt{3}a/2)$, $\vec{\delta}_2 = (a/2, -\sqrt{3}a/2)$, and $\vec{\delta}_3 = (-a, 0)$. The solid (hollow) circle represent sublattice A (B). (b) The Brillouin zone of the honeycomb lattice.

Introducing the transform

$$\hat{c}_{i\sigma}^{\dagger} = \frac{1}{\sqrt{N_\Lambda}} \sum_{\vec{k}} \hat{c}_{\vec{k}\sigma}^{\dagger} e^{-i\vec{k}\cdot\vec{R}_i} , \tag{3}$$

we can obtain the Hamiltonian in momentum space as

$$H_{DKM}=\sum_{\vec{k}}\psi_{\vec{k}}^{\dagger}H_{\vec{k}}\psi_{\vec{k}}\ . \tag{4}$$

Where $N_{\Lambda}$ is the number of unit cells, $\psi_{\vec{k}}=\left(\hat{c}_{\vec{k}\uparrow}^{A},\hat{c}_{\vec{k}\uparrow}^{B},\hat{c}_{\vec{k}\downarrow}^{A},\hat{c}_{\vec{k}\downarrow}^{B}\right)^{T}$ is the electron operators in momentum space, and the Bloch Hamiltonian

$$H_{\vec{k}}=\begin{pmatrix} \lambda\gamma(\vec{k}) & -g_{\alpha}(\vec{k}) & 0 & 0 \\ -g_{\alpha}^{*}(\vec{k}) & -\lambda\gamma(\vec{k}) & 0 & 0 \\ 0 & 0 & -\lambda\gamma(\vec{k}) & -g_{\alpha}(\vec{k}) \\ 0 & 0 & -g_{\alpha}^{*}(\vec{k}) & \lambda\gamma(\vec{k}) \end{pmatrix}. \tag{5}$$

where $g_{\alpha}(\vec{k})=t(\alpha e^{i\vec{k}\cdot\vec{\delta}_1}+e^{i\vec{k}\cdot\vec{\delta}_2}+e^{i\vec{k}\cdot\vec{\delta}_3})\quad =t(\alpha e^{i(k_x/2+\sqrt{3}k_y/2)}+e^{i(k_x/2-\sqrt{3}k_y/2)}+e^{-ik_x})$ and $\gamma=2[\sin(\vec{k}\cdot\vec{a}_1)-\sin(\vec{k}\cdot\vec{a}_2)-\sin(\vec{k}\cdot\vec{a}_3)]\quad =2[-\sin(\sqrt{3}k_y)+2\cos(3k_x/2)\sin(\sqrt{3}k_y/2)]$ ($a=1$).The single particle spectrum is obtained by diagonalizing the Hamiltonian as

$$E_{\pm}(k)=\pm\sqrt{\left|g_{\alpha}(\vec{k})\right|^{2}+\left(\lambda\gamma(\vec{k})\right)^{2}}\ . \tag{6}$$

Fig.2. Gaps with the change of $\alpha$ at various value of $\lambda$. It is obvious that the value of $\alpha$ at which the gap closes is independent on $\lambda$.

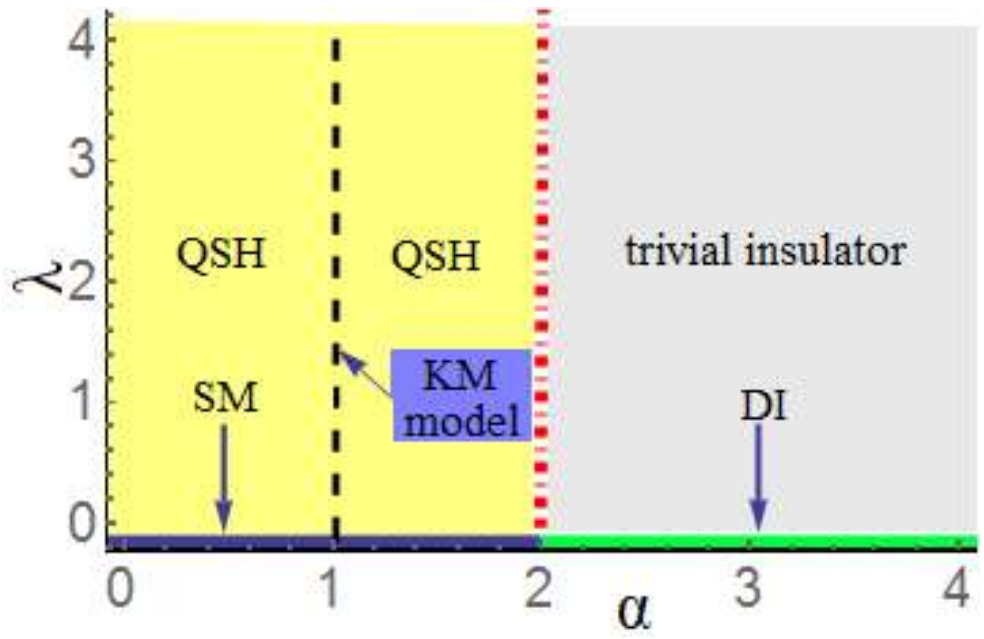

Fig.3. The phase diagram of the dimerized Kane-Mele model, including four phases: quantum spin Hall (QSH) state, semi-metallic (SM) state, dimerized insulator (DI), and trivial insulator. The dash line shows the QSH state possessed by the Kane-Mele model.

Several phases can be identified from the spectrum. At $\alpha = 1$ and $\lambda \neq 0$, the model is actually the Kane-Mele model. The system is a $Z_2$ topological insulator, i.e. the QSH state when electrons are half-filling and the chemical potential lies in the gap. In other regions of the $\alpha - \lambda$ space, phases connect adiabatically to the QSH state as long as the gap of the spectrum does not close and reopen and the system keep in the QSH state. For some values of $\alpha$ and $\lambda$, the gap close (then reopen) and the topological phase transition occurs. We find that the gap closes (then opens again) at $\alpha = 2$, and is independent on the value of $\lambda$. Gaps for various values of $\lambda$ are shown in Fig.2. When $\alpha > 2$ and $\lambda \neq 0$, the gap opens again and the phase is a trivial topologically insulator. Actually, at small $\alpha$ and $\lambda = 0$, the phase is a semi-metallic state that connected adiabatically to the one of graphene. There is a phase transition that semi-metallic state changes to dimerized insulator at $\alpha = 2$. The completed phase diagram of the dimerized Kane-Mele model is shown in Fig.3.

## 3. The dimerized Kane-Mele-Hubbard model

In this section, we add an on-site Hubbard term in the generalized Kane-Mele model to describe the strong interaction between electrons. The Hubbard model [31] is the archetypal model in the study about many body systems with strong interactions between electrons. It may be the simplest possible model which captures the essential physics of strongly correlated systems, e.g. metal-insulator transitions. It is well known that there are two descriptions of metal-insulator transitions, i.e. the Mott scenario [31, 32] and the magnetic scenario [33]. Here, we obtain boundaries of the Mott transition and the magnetic transition respectively in the dimerized Kane-Mele-Hubbard model and investigate effects of hopping anisotropy on transitions using different methods.

### 3.1. The Mott scenario of transitions

In this subsection, we investigate the interplay of the topology, the strong interaction, and the hopping anisotropy using the slave-rotor mean field theory [29].

#### 3.1.1. Slave-rotor representation

The Hamiltonian of the dimerized Kane-Mele-Hubbard model is

$$H_{DKMH} = H_{DKM} + H_U \tag{7}$$

and $H_U$, the on-site Hubbard interaction, reads

$$H_U = \frac{U}{2}\sum_i \left( \sum_\sigma n_{i\sigma} - 1 \right)^2 . \quad (8)$$

The slave-rotor representation of the electron annihilation operator is

$$\hat{c}_{i\sigma} = e^{i\theta_i} \hat{f}_{i\sigma} . \quad (9)$$

Here $e^{i\theta_i}$ is the U(1) rotor operator that describes the charge degree of freedom of the electron, and $\hat{f}_{i\sigma}$ is the spinon operator that describes the spin degree of freedom of the electron. Like Florens and Georges did, we introduce the canonical angular momentum $\hat{L} = -i\partial_\theta$ associated with the angular $\theta$. To recover the Hilbert space of the electron, the charge and spin degree of freedom should be satisfied the constraint

$$\sum_\sigma \hat{f}^\dagger_{i\sigma} \hat{f}_{i\sigma} + \hat{L}_i = 1 . \quad (10)$$

It is worthy to note that the constraint is different from the one introduced by Florens and Georges [29]. There are no minus in the exponent in the slave-rotor representation of the electron annihilation operator in Eq. (9). In the new operator presentation, the Hamiltonian reads

$$H_{DKMH} = -\alpha t \sum_{<i,j>\delta_1} \sum_\sigma e^{-i\theta_{ij}} \hat{f}^\dagger_{i\sigma} \hat{f}_{j\sigma} - t \sum_{<i,j>\delta_{2,3}} \sum_\sigma e^{-i\theta_{ij}} \hat{f}^\dagger_{i\sigma} \hat{f}_{j\sigma}$$
$$+ i\lambda \sum_{<<i,j>>} \sum_{\sigma\sigma'} v_{ij} e^{-i\theta_{ij}} \hat{f}^\dagger_{i\sigma} \sigma^z_{\sigma\sigma'} \hat{f}_{j\sigma'} + \frac{U}{2}\sum_i \hat{L}_i^2 - \mu \sum_i \sum_\sigma \hat{f}^\dagger_{i\sigma} \hat{f}_{i\sigma} . \quad (11)$$

Here $\theta_{ij} = \theta_i - \theta_j$ and $\mu$ is the chemical potential. The action of the system is

$$S = \int_0^\beta d\tau (-iL_i \partial_\tau \theta_i + f_i^* \partial_\tau f_i + H_{DKMH}) . \quad (12)$$

Here the imaginary time $\tau = it$ , and $L_i = -i\partial_{\theta_i} = (i/U)\partial_\tau \theta_i$ . The constraint has to be fulfilled via introducing the Lagrange multiplier $h_i$ to the action. The Lagrange multiplier is a constant $h$ for all sites at mean field level. Then, the action reads as

$$S = \int_0^\beta d\tau \Bigg[ \frac{1}{2U} \sum_i (\partial_\tau \theta_i + ih)^2 + \sum_i \sum_\sigma f^*_{i\sigma} (\partial_\tau - \mu + h) f_{i\sigma} + \sum_i (-h + \frac{h^2}{2U})$$
$$- \alpha t \sum_{<i,j>\delta_1} \sum_\sigma e^{-i\theta_{ij}} f^*_{i\sigma} f_{j\sigma} - t \sum_{<i,j>\delta_{2,3}} \sum_\sigma e^{-i\theta_{ij}} f^*_{i\sigma} f_{j\sigma}$$
$$+ i\lambda \sum_{<<i,j>>} \sum_{\sigma\sigma'} v_{ij} e^{-i\theta_{ij}} f^*_{i\sigma} \sigma^z_{\sigma\sigma'} f_{j\sigma'} \Bigg] . \quad (13)$$

Introducing the X-fields which defined by $X = e^{i\theta}$, we obtain the action

$$S = \int_0^\beta d\tau \left[ \frac{1}{2U} \sum_i [(i\partial_\tau + ih)X_i^*][(i\partial_\tau + ih)X_i] + \sum_i \sum_\sigma f_{i\sigma}^* (\partial_\tau - \mu + h) f_{i\sigma} \right.$$
$$- \alpha t \sum_{<i,j>\delta_1} \sum_\sigma X_i^* X_j f_{i\sigma}^* f_{j\sigma} - t \sum_{<i,j>\delta_{2,3}} \sum_\sigma X_i^* X_j f_{i\sigma}^* f_{j\sigma}$$
$$\left. + i\lambda \sum_{<<i,j>>} \sum_{\sigma\sigma'} \nu_{ij} X_i^* X_j f_{i\sigma}^* \sigma_{\sigma\sigma'}^z f_{j\sigma'} + \sum_i \rho_i |X_i|^2 + \sum_i (-h + \frac{h^2}{2U}) \right]. \tag{14}$$

Here $\rho_i$ is the Lagrange multiplier for constraint $|X_i| = 1$. The above action can be simplified by the Hartree-Fock mean field decomposition as

$$S = \int_0^\beta d\tau \left[ \frac{1}{2U} \sum_i [(i\partial_\tau + ih)X_i^*][(i\partial_\tau + ih)X_i] + \sum_i \rho_i |X_i|^2 \right.$$
$$- \alpha t Q_X \sum_{<i,j>\delta_1} X_i^* X_j - t Q_X \sum_{<i,j>\delta_{2,3}} X_i^* X_j + \lambda Q_X' \sum_{<<i,j>>} X_i^* X_j$$
$$+ \sum_i \sum_\sigma f_{i\sigma}^* (\partial_\tau - \mu + h) f_{i\sigma}$$
$$- \alpha t Q_f \sum_{<i,j>\delta_1} \sum_\sigma f_{i\sigma}^* f_{j\sigma} - t Q_f \sum_{<i,j>\delta_{2,3}} \sum_\sigma f_{i\sigma}^* f_{j\sigma} + i\lambda Q_f' \sum_{<<i,j>>} \sum_{\sigma\sigma'} \nu_{ij} f_{i\sigma}^* \sigma_{\sigma\sigma'}^z f_{j\sigma'}$$
$$\left. + \sum_i (-h + \frac{h^2}{2U}) + \cdots \right]. \tag{15}$$

Here $$Q_X = \left\langle \sum_\sigma f_{i\sigma}^{A*} f_{j\sigma}^B \right\rangle_{<i,j>}, \tag{16}$$

$$Q_f = \left\langle e^{-i\theta_{ij}} \right\rangle_{<i,j>}, \tag{17}$$

$$Q_X' = \left\langle \sum_{\sigma\sigma'} i\nu_{ij} f_{i\sigma}^* \sigma_{\sigma\sigma'}^z f_{j\sigma'} \right\rangle_{<<i,j>>}, \tag{18}$$

$$Q_f' = \left\langle e^{-i\theta_{ij}} \right\rangle_{<<i,j>>}, \tag{19}$$

and the symbol "$\cdots$" denote constant terms of mean field decomposition. The action can be transformed into frequency-momentum space via Fourier transforms

$$X_i(\tau) = (1/\sqrt{\beta N_\Lambda}) \sum_{\vec{k}} \sum_n e^{i(\vec{k}\cdot\vec{R} - \nu_n \tau)} X_{\vec{k}}(i\nu_n) \tag{20}$$

and

$$f_{i\sigma}(\tau) = (1/\sqrt{\beta N_\Lambda}) \sum_{\vec{k}} \sum_n e^{i(\vec{k}\cdot\vec{R} - \omega_n \tau)} f_{\vec{k}\sigma}(i\omega_n). \tag{21}$$

as

$$S=\Bigg[\frac{1}{2U}\sum_{\vec{k},n}v_n^2(X_{\vec{k}}^{A*}X_{\vec{k}}^{A}+X_{\vec{k}}^{B*}X_{\vec{k}}^{B})+\rho\sum_{\vec{k}}(X_{\vec{k}}^{A*}X_{\vec{k}}^{A}+X_{\vec{k}}^{B*}X_{\vec{k}}^{B})$$

$$+Q_X\sum_{\vec{k}}\left(-g_\alpha(\vec{k})\right)X_{\vec{k}}^{A*}X_{\vec{k}}^{B}+Q_X\sum_{\vec{k}}\left(-g_\alpha^*(\vec{k})\right)X_{\vec{k}}^{B*}X_{\vec{k}}^{A}$$

$$+\lambda Q'_X\sum_{\vec{k}}\gamma_X(\vec{k})X_{\vec{k}}^{A*}X_{\vec{k}}^{A}+\lambda Q'_X\sum_{\vec{k}}\gamma_X(\vec{k})X_{\vec{k}}^{B*}X_{\vec{k}}^{B}$$

$$+\sum_{\vec{k},n}\sum_{\sigma}i\omega_n(f_{\vec{k}\sigma}^{A*}f_{\vec{k}\sigma}^{A}+f_{\vec{k}\sigma}^{B*}f_{\vec{k}\sigma}^{B})$$

$$+Q_f\sum_{\vec{k},n}\sum_{\sigma}\left(-g_\alpha(\vec{k})\right)f_{\vec{k}\sigma}^{A*}f_{\vec{k}\sigma}^{B}+Q_f\sum_{\vec{k},n}\sum_{\sigma}\left(-g_\alpha^*(\vec{k})\right)f_{\vec{k}\sigma}^{B*}f_{\vec{k}\sigma}^{A}$$

$$+\lambda Q'_f\sum_{\vec{k}}\sum_{\sigma\sigma'}\gamma_f(\vec{k})f_{\vec{k}\sigma}^{A*}\sigma_{\sigma\sigma'}^{z}f_{\vec{k}\sigma'}^{A}+\lambda Q'_f\sum_{\vec{k}}\left(-\gamma_f(\vec{k})\right)f_{\vec{k}\sigma}^{B*}\sigma_{\sigma\sigma'}^{z}f_{\vec{k}\sigma'}^{B}\Bigg]$$

$$+\sum_i(-h+\frac{h^2}{2U})+\cdots. \tag{22}$$

Here $\gamma_X(\vec{k})=2[\cos(\vec{k}\cdot\vec{a}_1)+\cos(\vec{k}\cdot\vec{a}_2)+\cos(\vec{k}\cdot\vec{a}_3)$

$$=2[\cos(3k_x/2+\sqrt{3}k_y/2)+\cos(3k_x/2-\sqrt{3}k_y/2)+\cos(\sqrt{3}k_y),$$

$$\gamma_f(\vec{k})=2[\sin(\vec{k}\cdot\vec{a}_1)-\sin(\vec{k}\cdot\vec{a}_2)-\sin(\vec{k}\cdot\vec{a}_3)$$

$$=2[\sin(3k_x/2+\sqrt{3}k_y/2)-\sin(3k_x/2-\sqrt{3}k_y/2)-\sin(\sqrt{3}k_y),$$

and we set $\mu=-h=0$ for half-filled case. We simplify the action in matrix form as

$$S=\sum_{\vec{k},n}\psi_\eta^{X*}[(\frac{v_n^2}{2U}+\rho)\delta_{\eta\kappa}+h_{\eta\kappa}^{X}]\psi_\kappa^{X}+\sum_{\vec{k},n}\psi_\eta^{f*}[(i\omega_n)\delta_{\eta\kappa}+h_{\eta\kappa}^{f}]\psi_\kappa^{f}$$

$$+\sum_i(-h+\frac{h^2}{2U})+\cdots. \tag{23}$$

Here $\psi^X=(X_{\vec{k}}^A,X_{\vec{k}}^B)^T$, $\psi^f=(f_{\vec{k}\uparrow}^A,f_{\vec{k}\uparrow}^B,f_{\vec{k}\downarrow}^A,f_{\vec{k}\downarrow}^B)^T$. Hamiltonian matrices of the X-field and the spinon are respectively

$$h^X=\begin{pmatrix}Q'_X\lambda\gamma_X(\vec{k}) & -Q_Xg_\alpha(\vec{k})\\ -Q_Xg_\alpha^*(\vec{k}) & Q'_X\lambda\gamma_X(\vec{k})\end{pmatrix}. \tag{24}$$

and

$$h^f = \begin{pmatrix} Q'_f \lambda \gamma_f(\vec{k}) & -Q_f g_\alpha(\vec{k}) & 0 & 0 \\ -Q_f g_\alpha^*(\vec{k}) & -Q'_f \lambda \gamma_f(\vec{k}) & 0 & 0 \\ 0 & 0 & -Q'_f \lambda \gamma_f(\vec{k}) & -Q_f g_\alpha(\vec{k}) \\ 0 & 0 & -Q_f g_\alpha^*(\vec{k}) & Q'_f \lambda \gamma_f(\vec{k}) \end{pmatrix}. \tag{25}$$

Diagonalized Hamiltonians are obtained as

$$h^X = \begin{pmatrix} E_+^X(\vec{k}) & 0 \\ 0 & E_-^X(\vec{k}) \end{pmatrix} \tag{26}$$

and

$$h^f = \begin{pmatrix} E_+^f(\vec{k}) & 0 & 0 & 0 \\ 0 & E_-^f(\vec{k}) & 0 & 0 \\ 0 & 0 & E_+^f(\vec{k}) & 0 \\ 0 & 0 & 0 & E_-^f(\vec{k}) \end{pmatrix}. \tag{27}$$

Here $E_\pm^X(\vec{k}) = \pm Q_X \left| g_\alpha(\vec{k}) \right| + Q'_X \lambda \gamma_X(\vec{k})$ and

$E_\pm^f(\vec{k}) = \pm\sqrt{\left(Q_f \left| g_\alpha(\vec{k}) \right|\right)^2 + \left(Q'_f \lambda \gamma_f(\vec{k})\right)^2}$ . Diagonalizing the two Hamiltonians, the similar matrices corresponding to Hamiltonians of X-field and spinon are respectively

$$U_X = \frac{1}{\sqrt{2}} \begin{pmatrix} -\dfrac{b_X}{|b_X|} & \dfrac{b_X}{|b_X|} \\ 1 & 1 \end{pmatrix} \tag{28}$$

and

$$U_{\uparrow(\downarrow)} = \frac{1}{\sqrt{2}} \begin{pmatrix} -\dfrac{b_f}{N_\mp} & \dfrac{b_f}{N_\pm} \\ \dfrac{\sqrt{a_f^2 + |b_f|^2} \mp a_f}{N_\mp} & \dfrac{\sqrt{a_f^2 + |b_f|^2} \pm a_f}{N_\pm} \end{pmatrix}. \tag{29}$$

Here $N_\pm = \sqrt{2\sqrt{a_f^2 + |b_f|^2}\left(\sqrt{a_f^2 + |b_f|^2} \pm a_f\right)}$, $a_f = Q'_f \lambda \gamma_f(\vec{k})$, $b_X = Q_X g_\alpha(\vec{k})$, and

$b_f = Q_f g_\alpha(\vec{k})$.

Finally, Green's functions of the X-field and the spinon in the lower band can be obtained as

$$G_X^l(\vec{k},iv_n)=\frac{1}{v_n^2/2U+\rho+E_-^X(\vec{k})}, \tag{30}$$

and

$$G_f^l(\vec{k},i\omega_n)=\frac{1}{i\omega_n+E_-^f(\vec{k})}. \tag{31}$$

3.1.2. The self-consistency equations

For the X-field, the constrain equation is $|X_i(\tau)|^2=1$ or $\sum_i\left\langle X_i^*(\tau)X_i(\tau)\right\rangle/2N_\Lambda=1$. It is satisfied on average for all sites, i.e.

$$\frac{1}{N_\Lambda}\sum_{\vec{k}}\frac{1}{\beta}\sum_n G_X^l(\vec{k},iv_n)=1. \tag{32}$$

Carrying out Matsubara sum over the frequencies, we can obtain the equation as

$$\frac{1}{N_\Lambda}\sum_{\vec{k}}\frac{U}{2\sqrt{U(\rho+E_-^X(\vec{k}))}}=1. \tag{33}$$

When the Mott transition occurs, the gap of the X-field is closed, i.e.

$$\rho=-\min(E_-^X(\vec{k})). \tag{34}$$

Finally, we obtain the first self-consistency equation

$$\frac{1}{N_\Lambda}\sum_{\vec{k}}\frac{U_c}{2\sqrt{U_c(E_-^X(\vec{k})-\min(E_-^X(\vec{k}))}}=1. \tag{35}$$

Here $U_c$ is the critical value of the Hubbard interaction at which the Mott transition occurs.

The second self-consistency equation is actually Eq. (16)

$$Q_X=\left\langle\sum_\sigma f_{i\sigma}^{A*}f_{j\sigma}^B\right\rangle_{<i,j>}$$

$$=\frac{1}{3N_\Lambda}\sum_{<i,j>}\sum_\sigma\left\langle f_{i\sigma}^{A*}f_{j\sigma}^B\right\rangle_{<i,j>}$$

$$=\frac{1}{3N_\Lambda}\sum_{\vec{k}}\sum_{\sigma}\frac{g_\alpha(\vec{k})}{t}\left\langle f_{\vec{k}\sigma}^{A*}f_{\vec{k}\sigma}^{B}\right\rangle$$

$$=\frac{1}{3N_\Lambda t}\sum_{\vec{k}}\frac{Q_f\left|g_\alpha(\vec{k})\right|^2}{\sqrt{(Q_f'\lambda\gamma_f(\vec{k}))^2+(Q_f\left|g_\alpha(\vec{k})\right|)^2}}\,. \tag{36}$$

The third self-consistency equation is

$$Q_X'=\left\langle\sum_{\sigma\sigma'}iv_{ij}f_{i\sigma}^*\sigma_{\sigma\sigma'}^z f_{j\sigma'}\right\rangle_{<<i,j>>}$$

$$=\frac{1}{6N_\Lambda}\sum_{<<i,j>>}\left(\sum_{\sigma\sigma'}\left\langle iv_{ij}f_{i\sigma}^{A*}\sigma_{\sigma\sigma'}^z f_{j\sigma'}^{A}\right\rangle+(i\leftrightarrow j)\right)$$

$$=\frac{1}{6N_\Lambda}\sum_{\vec{k}}\sum_{\sigma\sigma'}\gamma_f(\vec{k})\sigma_{\sigma\sigma'}^z\left\langle f_{i\sigma}^{A*}f_{j\sigma'}^{A}\right\rangle$$

$$=\frac{1}{6N_\Lambda}\sum_{\vec{k}}\frac{-Q_f'\lambda\gamma_f^2(\vec{k})}{\sqrt{(Q_f'\lambda\gamma_f(\vec{k}))^2+(Q_f\left|g_\alpha(\vec{k})\right|)^2}}\,. \tag{37}$$

The fourth self-consistency equation is

$$Q_f=\left\langle X_i^{A*}X_j^{B}\right\rangle_{<i,j>}$$

$$=\frac{1}{3N_\Lambda}\sum_{<i,j>}\left\langle X_i^{A*}X_j^{B}\right\rangle$$

$$=\frac{1}{3N_\Lambda}\sum_{\vec{k}}\frac{g_\alpha(\vec{k})}{t}\left\langle X_{\vec{k}}^{A*}X_{\vec{k}}^{B}\right\rangle$$

$$=\frac{1}{12N_\Lambda}\sum_{\vec{k}}\frac{\left|g_\alpha(\vec{k})\right|}{t}\frac{U}{\sqrt{U(\rho+E_-^X(\vec{k}))}}\,. \tag{38}$$

The fifth self-consistency equation is

$$Q_f'=\left\langle X_i^*X_j\right\rangle_{<<i,j>>}$$

$$=\frac{1}{6N_{\Lambda}}\sum_{<<i,j>>}\left(\left\langle X_i^{A*}X_j^{A}\right\rangle+(i\leftrightarrow j)\right)$$

$$=\frac{1}{6N_{\Lambda}}\sum_{\vec{k}}\gamma_X(\vec{k})\left\langle X_{\vec{k}}^{A*}X_{\vec{k}}^{A}\right\rangle$$

$$=\frac{1}{24N_{\Lambda}}\sum_{\vec{k}}\gamma_X(\vec{k})\frac{U}{\sqrt{U(\rho+E_-^X(\vec{k}))}}. \quad (39)$$

3.1.3. Results

From Eq. (11), we can see that the band of the spinon of the dimerized Kane-Mele-Hubbard model has the same topological structure as the electron of the dimerized Kane-Mele model. When rotors (charge sector) are condensed, they combine spinons to form electrons with the non-trivial topological band structure and the phase is a QSH state or topological band insulator (TBI). When rotors are uncondensed at some large Hubbard interaction U, the charge sector is a Mott insulator, while the spinon sector may have the non-trivial topological band structure. The phase is an exotic topological phase named topological Mott insulator (TMI) [14]. At the larger U, the magnetic transition occurs and destroys the topological band structure of spinons. it will be discussed in the following subsection. Solving numerically the five self-consistency equations, i.e. Eq. (35)-Eq. (39), we obtain boundaries of the Mott transition at different $\alpha$ as shown in Fig.4.

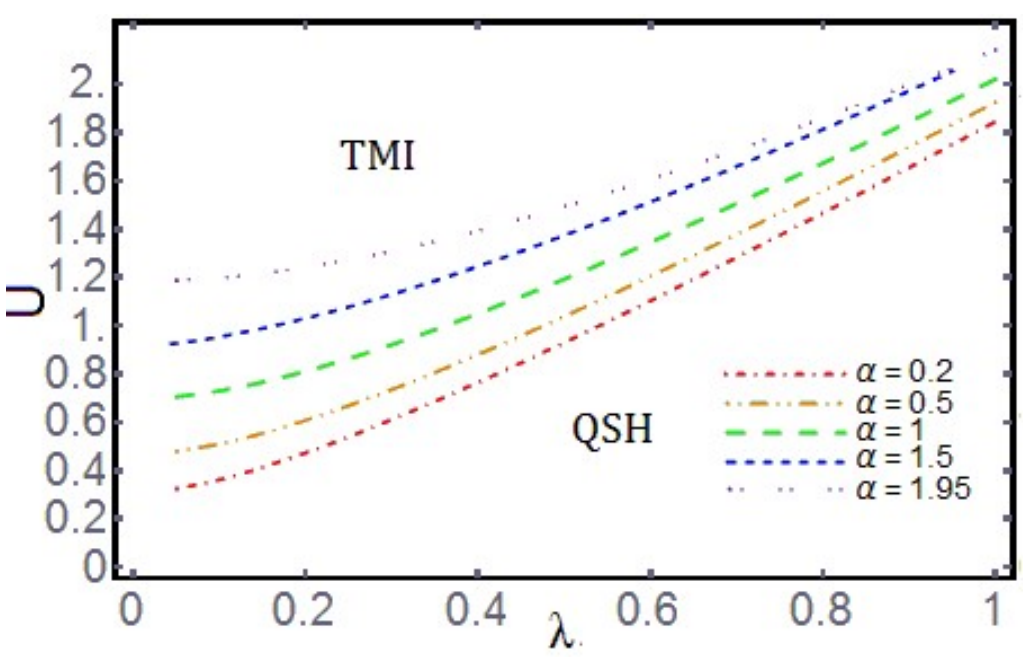

Fig.4. Boundaries of the Mott transition of charge sector at various values of $\alpha$

For each $\alpha$, the phase in the upper left region is the TMI and there is a spin-charge separation. The charge is frozen in this Mott insulating state and the spinon has the same band structure as the electron of the DKM model. More precisely, spinons form the U(1) spin liquid state [19]. The phase in the lower right region is the QSH state (TBI). In this region, the DKMH model is actually the renormalized DKM model and the spinon combines the condensed charge to form the electron with non-trivial topological band structure. From results of numerical calculation, we can find that the larger Hubbard interaction U is needed to destroy the condensation of charges (i.e. destroy the QSH state) if the spin-orbital coupling $\lambda$ becomes large, and the larger U is also needed to destroy the condensation of charges if the value of $\alpha$ becomes larger.

The above discussion is based on the mean-field approximation. When fluctuations are take into account, the U(1) spin liquid of spinons must support an emergent dynamical U(1) gauge field. It is not clear that whether the gauge field can destabilizes the spin liquid state of spinons, i.e. the TMI in our model. In fact, another layer is needed to screen the gauge field and suppress the fluctuation [19]. In this work, we do not concern fluctuations and assume that the TMI is stable

3.2 The magnetic scenario of transition

In this subsection, we obtain the boundary of the magnetic transition using Hartree-Fock mean-field theory. The topological band structure of the spinon should be destroyed, and TMI transforms to the spin density wave (SDW) state.

The Hubbard interaction reads

$$\begin{aligned} H_U &= U\sum_i n_{i\uparrow}n_{i\downarrow} \\ &= \frac{U}{4}\sum_i [(n_{i\uparrow}+n_{i\downarrow})^2-(n_{i\uparrow}-n_{i\downarrow})^2]. \end{aligned} \tag{40}$$

The interaction can be wrote in Hartree-Fock mean-field decomposition

$$\begin{aligned} H_U^{HF} &= \frac{U}{2}\sum_i [-m^A(n_{i\uparrow}^A-n_{i\downarrow}^A)-m^B(n_{i\uparrow}^B-n_{i\downarrow}^B)] \\ &\quad +\frac{U}{4}\sum_i [(n_i^A)^2+(n_i^B)^2]+\frac{N_\Lambda U}{4}[(m^A)^2+(m^B)^2]. \end{aligned} \tag{41}$$

Here $n_i = n_{i\uparrow}+n_{i\downarrow}$ is the number of electron at site $i$, $m_i = n_{i\uparrow}-n_{i\downarrow}$ is the magnetic mean-field parameter, and the sum over all of primitive cells. For simplification, we set $m^A=-m^B=m$ and obtain the Hubbard interaction in momentum space as

$$H_U^{HF} = \frac{Um}{2}\sum_{\vec{k}} [-n_{\vec{k}\uparrow}^A+n_{\vec{k}\downarrow}^A+n_{\vec{k}\uparrow}^B-n_{\vec{k}\downarrow}^B)]+\frac{UN_\Lambda}{2}m^2+\frac{U}{4}[(n_i^A)^2+(n_i^B)^2]. \tag{42}$$

The full Hamiltonian of the DKMH model can be obtained finally as

$$H_{DKMH}^{HF} = \sum_{\vec{k}} \psi_{\vec{k}}^{\dagger} h_{\vec{k}}^{HF} \psi_{\vec{k}} + \frac{UN_{\Lambda}}{2} m^2 + \frac{U}{4}[(n_i^A)^2 + (n_i^B)^2]. \tag{43}$$

Here $\psi_{\vec{k}} = (c_{\vec{k}\uparrow}^A, c_{\vec{k}\downarrow}^A, c_{\vec{k}\uparrow}^B, c_{\vec{k}\downarrow}^B)^T$,

and

$$h_{\vec{k}}^{HF} = \begin{pmatrix} \lambda\gamma(\vec{k}) - \frac{Um}{2} & -g_{\alpha}(\vec{k}) & 0 & 0 \\ -g_{\alpha}^*(\vec{k}) & -(\lambda\gamma(\vec{k}) - \frac{Um}{2}) & 0 & 0 \\ 0 & 0 & -(\lambda\gamma(\vec{k}) - \frac{Um}{2}) & -g_{\alpha}(\vec{k}) \\ 0 & 0 & -g_{\alpha}^*(\vec{k}) & \lambda\gamma(\vec{k}) - \frac{Um}{2} \end{pmatrix}.$$

Diagonalizing the Hamiltonian, we can get the free energy

$$F(m) = -2\sum_{\vec{k}} \sqrt{\left|g_{\alpha}(\vec{k})\right|^2 + \left(\lambda\gamma(\vec{k}) - Um/2\right)^2} + const. \tag{44}$$

Minimizing the free energy with respect to m yield the self-consistency equation

$$m = \frac{1}{N_{\Lambda}} \sum_{\vec{k}} \frac{Um/2 - \lambda\gamma(\vec{k})}{\sqrt{\left|g_{\alpha}(\vec{k})\right|^2 + \left(\lambda\gamma(\vec{k}) - Um/2\right)^2}}. \tag{45}$$

We solve numerically Eq. (45) and obtain boundaries of magnetic transitions as shown in Fig.5.

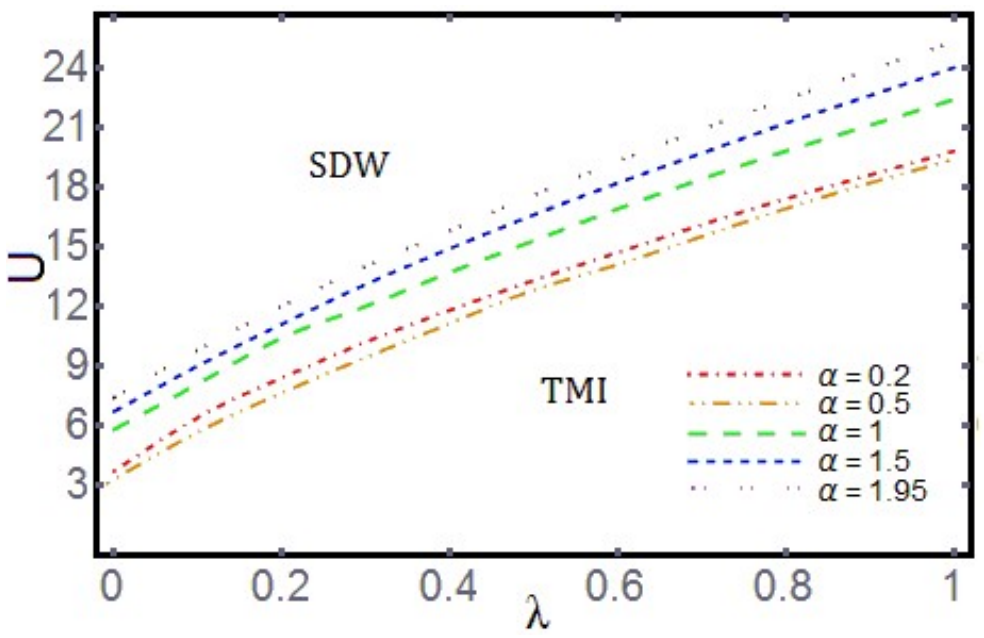

Fig.5. Boundaries of the magnetic transition of the DKMH model as obtained from Hartree-Fock mean field theory at various values of $\alpha$.

The influence of $\alpha$ on the boundary of the magnetic transition is same as the situation about the Mott transition of the charge sector. At the lager $\alpha$, the lager $U$ is needed to destroy the topological phase. For clarity, completed phase diagrams including the magnetic transition and the Mott transition at some values of $\alpha$ are shown in Fig. 6.

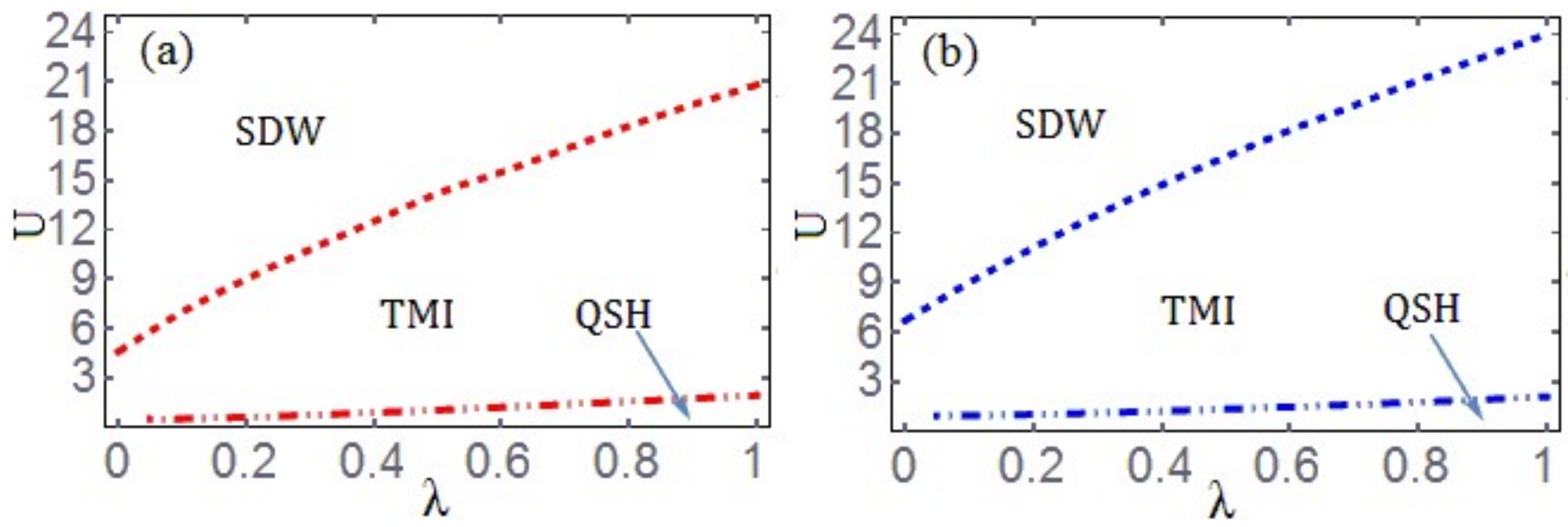

Fig.6. Completed phase diagrams including the Mott transition and the magnetic transition at $\alpha = 0.5$ (a) and $\alpha = 1.5$ (b).

## 4. Conclusions

We have investigated the dimerized Kane-Mele model without interactions and got a completed phase diagram including the QSH state corresponding to the Kane-Mele model. The most important result is that the transition from QSH state to trivial topologically insulator is dependent on parameter $\alpha$ only. According to our investigation, the topological transition occurs at $\alpha = 2$.

Then, we have studied a correlated dimerized Kane-Mele model with the Hubbard interaction. We focused on two different transitions, i.e. the Mott transition and the magnetic transition of the dimerized Kane-Mele-Hubbard model and obtained boundaries of the two transitions at various values of $\alpha$. For the Mott transition, using the slave-rotors mean field method, we obtained the spin-charge separation and investigated the condensation of the charge. When the charge uncondensed, the charge sector is an insulating phase, while the spin sector has the non-trivial topological band structure. The phase is a topological Mott insulator. The spin sector possesses the spin liquid state actually. The quantum fluctuation is important to spin liquids state and to the stability of the TMI. However, benefiting from the discussion by Yong *et al* [19], we did not concern it in this work. Effects of quantum fluctuations on the DKMH model will discussed in the future. For the Magnetic transition, using the Hartree-Fock mean field method, we obtained the boundary of the transition from the TMI to the SDW state. We find that changes of boundaries of the two transitions have the same trend. The critical Hubbard interaction increase with the increasing of the spin-orbital coupling $\lambda$ at same value of $\alpha$. The critical Hubbard interaction increase with the increasing of the parameter $\alpha$ at same spin-orbital coupling $\lambda$. In the end of this work, we show respectively the completed phase diagrams at $\alpha = 0.5$ and $\alpha = 1.5$ for integrity.

## 5. Acknowledgements

This work was supported by Applied Basic Research Program of Yunnan Provincial Science and Technology Department under grant Nos.2013FZ083, 2013FZ028 and Foundation of Yunnan Educational Committee under the grant No. 2012Z040. TD and YXL also acknowledge support from NSFC under grant No. 61640415. YL acknowledges support from NSFC under grant No.61461055.